\documentstyle[12pt,epsf,psfig]{article}
\begin{document}

\baselineskip=7mm
\def\ap#1#2#3{           {\it Ann. Phys. (NY) }{\bf #1} (19#2) #3}
\def\arnps#1#2#3{        {\it Ann. Rev. Nucl. Part. Sci. }{\bf #1} (19#2) #3}
\def\cnpp#1#2#3{        {\it Comm. Nucl. Part. Phys. }{\bf #1} (19#2) #3}
\def\apj#1#2#3{          {\it Astrophys. J. }{\bf #1} (19#2) #3}
\def\asr#1#2#3{          {\it Astrophys. Space Rev. }{\bf #1} (19#2) #3}
\def\ass#1#2#3{          {\it Astrophys. Space Sci. }{\bf #1} (19#2) #3}

\def\apjl#1#2#3{         {\it Astrophys. J. Lett. }{\bf #1} (19#2) #3}
\def\ass#1#2#3{          {\it Astrophys. Space Sci. }{\bf #1} (19#2) #3}
\def\jel#1#2#3{         {\it Journal Europhys. Lett. }{\bf #1} (19#2) #3}

\def\ib#1#2#3{           {\it ibid. }{\bf #1} (19#2) #3}
\def\nat#1#2#3{          {\it Nature }{\bf #1} (19#2) #3}
\def\nps#1#2#3{          {\it Nucl. Phys. B (Proc. Suppl.) } {\bf #1} (19#2) #3}
\def\np#1#2#3{           {\it Nucl. Phys. }{\bf #1} (19#2) #3}

\def\pl#1#2#3{           {\it Phys. Lett. }{\bf #1} (19#2) #3}
\def\pr#1#2#3{           {\it Phys. Rev. }{\bf #1} (19#2) #3}
\def\prep#1#2#3{         {\it Phys. Rep. }{\bf #1} (19#2) #3}
\def\prl#1#2#3{          {\it Phys. Rev. Lett. }{\bf #1} (19#2) #3}
\def\pw#1#2#3{          {\it Particle World }{\bf #1} (19#2) #3}
\def\ptp#1#2#3{          {\it Prog. Theor. Phys. }{\bf #1} (19#2) #3}
\def\jppnp#1#2#3{         {\it J. Prog. Part. Nucl. Phys. }{\bf #1} (19#2) #3}

\def\rpp#1#2#3{         {\it Rep. on Prog. in Phys. }{\bf #1} (19#2) #3}
\def\ptps#1#2#3{         {\it Prog. Theor. Phys. Suppl. }{\bf #1} (19#2) #3}
\def\rmp#1#2#3{          {\it Rev. Mod. Phys. }{\bf #1} (19#2) #3}
\def\zp#1#2#3{           {\it Zeit. fur Physik }{\bf #1} (19#2) #3}
\def\fp#1#2#3{           {\it Fortschr. Phys. }{\bf #1} (19#2) #3}
\def\Zp#1#2#3{           {\it Z. Physik }{\bf #1} (19#2) #3}
\def\Sci#1#2#3{          {\it Science }{\bf #1} (19#2) #3}

\def\n.c.#1#2#3{         {\it Nuovo Cim. }{\bf #1} (19#2) #3}
\def\r.n.c.#1#2#3{       {\it Riv. del Nuovo Cim. }{\bf #1} (19#2) #3}
\def\sjnp#1#2#3{         {\it Sov. J. Nucl. Phys. }{\bf #1} (19#2) #3}
\def\yf#1#2#3{           {\it Yad. Fiz. }{\bf #1} (19#2) #3}
\def\zetf#1#2#3{         {\it Z. Eksp. Teor. Fiz. }{\bf #1} (19#2) #3}
\def\zetfpr#1#2#3{         {\it Z. Eksp. Teor. Fiz. Pisma. Red. }{\bf #1} (19#2) #3}
\def\jetp#1#2#3{         {\it JETP }{\bf #1} (19#2) #3}
\def\mpl#1#2#3{          {\it Mod. Phys. Lett. }{\bf #1} (19#2) #3}
\def\ufn#1#2#3{          {\it Usp. Fiz. Naut. }{\bf #1} (19#2) #3}
\def\sp#1#2#3{           {\it Sov. Phys.-Usp.}{\bf #1} (19#2) #3}
\def\ppnp#1#2#3{           {\it Prog. Part. Nucl. Phys. }{\bf #1} (19#2) #3}
\def\cnpp#1#2#3{           {\it Comm. Nucl. Part. Phys. }{\bf #1} (19#2) #3}
\def\ijmp#1#2#3{           {\it Int. J. Mod. Phys. }{\bf #1} (19#2) #3}
\def\ic#1#2#3{           {\it Investigaci\'on y Ciencia }{\bf #1} (19#2) #3}
\def\tp{these proceedings}
\def\pc{private communication}
\def\ip{in preparation}
\newcommand{\TeV}{\,{\rm TeV}}
\newcommand{\GeV}{\,{\rm GeV}}
\newcommand{\MeV}{\,{\rm MeV}}
\newcommand{\keV}{\,{\rm keV}}
\newcommand{\eV}{\,{\rm eV}}
\newcommand{\Tr}{{\rm Tr}\!}
\renewcommand{\arraystretch}{1.2}
\newcommand{\be}{\begin{equation}}
\newcommand{\ee}{\end{equation}}
\newcommand{\beqa}{\begin{eqnarray}}
\newcommand{\eeqa}{\end{eqnarray}}
\newcommand{\ba}{\begin{array}}
\newcommand{\ea}{\end{array}}
\newcommand{\bmat}{\left(\ba}
\newcommand{\emat}{\ea\right)}
\newcommand{\refs}[1]{(\ref{#1})}
\newcommand{\ler}{\stackrel{\scriptstyle <}{\scriptstyle\sim}}
\newcommand{\ger}{\stackrel{\scriptstyle >}{\scriptstyle\sim}}
\newcommand{\lag}{\langle}
\newcommand{\rag}{\rangle}
\newcommand{\ns}{\normalsize}
\newcommand{\cm}{{\cal M}}
\newcommand{\gr}{m_{3/2}}
\newcommand{\p}{\partial}
\renewcommand{\le}{\left(}
\newcommand{\ri}{\right)}
\relax
\def\321{$SU(3)\times SU(2)\times U(1)$}
\def\ord{{\cal O}}
\def\tl{{\tilde{l}}}
\def\tL{{\tilde{L}}}
\def\bd{{\overline{d}}}
\def\tL{{\tilde{L}}}
\def\a{\alpha}
\def\b{\beta}
\def\g{\gamma}
\def\c{\chi}
\def\d{\delta}
\def\D{\Delta}
\def\db{{\overline{\delta}}}
\def\Db{{\overline{\Delta}}}
\def\e{\epsilon}
\def\l{\lambda}
\def\n{\nu}
\def\m{\mu}
\def\nt{{\tilde{\nu}}}
\def\p{\phi}
\def\P{\Phi}
\def\sol{\Delta_{\odot}}
\def\ue3{|U_{e3}|}
\def\sa{\theta_{\odot}}
\def\atm{\Delta_{\mathbf{atm}}}
\def\k{\kappa}
\def\x{\xi}
\def\r{\rho}
\def\s{\sigma}
\def\t{\tau}
\def\th{\theta}
\def\om{\omega}
\def\ne{\nu_e}
\def\nm{\nu_{\mu}}
\def\snui{\tilde{\nu_i}}
\def\ehat{\hat{e}}
\def\la{{\makebox{\tiny{\bf loop}}}}
\def\ta{\tilde{a}}
\def\tb{\tilde{b}}
\def\mb{m_{1b}}
\def\mt{m_{1 \tau}}
\def\rl{{\rho}_l}
\def\meg{\m \rightarrow e \g}

\renewcommand{\Huge}{\Large}
\renewcommand{\LARGE}{\Large}
\renewcommand{\Large}{\large}
\title{\large\bf Quark-Lepton Universality and Large Leptonic Mixing}
\author{ Anjan S. Joshipura$^1$ and A. Yu. Smirnov$^{2,3}$\\[.5cm]
{\ns\it  $^1$Theoretical Physics Group, Physical Research Laboratory,}\\
{\ns\it Navrangpura, Ahmedabad 380 009, India}\\
{\ns\it $^2$International Centre
for Theoretical Physics,}\\ {\ns \it Strada Costiera 11, 31014 Trieste, Italy}\\
{\ns\it $^3$Institute for Nuclear Research, Russian Academy of
Sciences, Moscow, Russia}}
\date{}

\maketitle
\begin{abstract}
A unified description of fermionic mixing is proposed which assumes that
in  certain basis ($i$) a single complex unitary matrix $V$ 
diagonalizes mass matrices of  all fermions to the leading order, ($ii$)   
the $SU(5)$ relation $M_d=M_l^T$  exists between the mass  matrices of the down 
quarks and the charged leptons, and ($iii$) $M_d^\dagger=M_d$. These assumptions 
automatically lead to 
different mixing patterns for quarks and leptons: quarks remain  unmixed 
to 
leading order ($i.~e.~ V_{CKM}^0=1$) while leptons have non-trivial  mixing given 
by a symmetric unitary matrix $V^0_{PMNS} = V^T V$. 
$V$ depends on two physical mixing angles and for values 
of these angles $\sim 20^\circ-25 ^\circ$ it reproduces the observed  
mixing patterns rather well. 
We identify conditions under which the universal mixing $V$  follows from the 
universal mass matrices of fermions. 
Relatively small perturbations to the leading order structure 
lead to the CKM mixing and corrections to 
$V^0_{PMNS}$. We find that if  the correction matrix  
equals the CKM matrix,  the resulting lepton mixing 
agrees well with data and predicts $\sin \theta_{13} > 0.08$. 
In a more general context, the  assumption of partial universality, ($i.e.$ different mixing for the 
up and the down components 
of doublets) is shown to lead to a complementarity  relation $V_{PMNS}V_{CKM}=V^TV$ in the lowest
order.
\end{abstract}
\newpage

\section{Introduction}

Experimental results show striking differences in the mixing
patterns of quarks and leptons: The quarks retain their flavour
approximately and mix very little; leptons mix strongly; two of the three
leptonic mixing angles are large and one of them may even be maximal.
Understanding these differences among quarks and leptons  
could be a key step to the theory of fermion masses and this 
problem is extensively discussed now \cite{rev}.

It may happen that quarks and leptons are fundamentally different: 
The large leptonic mixing or possible quasi-degenerate structure of the neutrino
mass spectrum  may be signals of some special    
symmetry operating in the leptonic sector alone. 
In particular, the maximal (or close to maximal) 
atmospheric mixing and vanishing 
(very small) $\sin\theta_{13}$ indicate presence  of symmetry like 
the $~\mu-\tau$ interchange symmetry~\cite{mt}. 
Degeneracy of spectrum and inverse mass hierarchy may be considered as an 
evidence of the   $L_e - L_\mu-L_\tau$ lepton number symmetry \cite{emt}.

Alternatively, the leptonic world may not be special and there may be 
underlying quark-lepton symmetry or unification. 
There are different ways to reconcile the quark-lepton unification 
with strongly different patterns of quark and lepton mixings.  
For example, the $SU(5)$ relation \cite{rev} 
\be 
\label{su5} 
M_l=M_d^T ~ 
\ee
between the down quark and the charged lepton mass matrices  can simultaneously 
accommodate small quark and large leptonic mixing angles if 
the matrices (\ref{su5}) have lopsided form \cite{lop1,lop2}. 
In $SO(10)$ the $b-\tau$
unification \cite{btau} can be related to large leptonic mixing if 
neutrino masses
dominantly come from  interactions with  the Higgs triplets 
\cite{seesaw2}.

The quark-lepton symmetry may take more general form than that stipulated in 
grand unified theories and attempts
to understand fermionic mixing along these lines postulate some
universal form for all fermionic mass matrices. The observed 
differences arise from small perturbations to this unified description.
Two examples discussed in the literature  are the democratic mass matrices 
\cite{demo} and singular structure with  hierarchical matrix elements 
\cite{sd}. First two generations of fermions remain massless in these ansatz 
and therefore small perturbations can make large differences in 
mass hierarchies and furthermore the seesaw mechanism with nearly singular 
mass matrix of the RH neutrinos lead to enhancement of lepton mixing \cite{sd}. 
While strictly maximal mixing is difficult to reconcile in this picture,
the large leptonic mixing can be reproduced in agreement with data. 

The quark-lepton unification may have the form of quark-lepton 
complementarity~\cite{qlc1,qlc2}: the data indicate that quark and lepton mixing angles 
(at least for the first and second generations)  sum up to maximal mixing, 
$\theta_{12}^l = \pi/4 - \theta_C$, where $\theta_C$ is the 
Cabibbo angle.  This last equality may imply that there is (i) some 
structure in the leptonic 
sector (only) which generates maximal (bi-maximal) mixing  and (ii)  
the quark-lepton symmetry which propagates  
the CKM - mixing from quarks to leptons. In view of strong difference of 
the mass hierarchies in 
the quark and lepton sectors the later implies that the CKM mixing 
matrix weakly depends on the mass hierarchies.

The aim of this paper is to discuss an alternative 
form of quark-lepton universality. This form is strongly motivated by grand unified 
theories,
particularly, eq. (\ref{su5}) and has the virtue that different patterns for 
quark and lepton mixing is built into this universal ansatz.
It is assumed that in certain (universality) basis  the relation between 
the basis states  and the mass eigenstates 
of all fermions is determined by  
a universal (complex) $3\times 3$ unitary matrix $V$ (or its conjugate).  
In addition, if eq. (\ref{su5}) is also imposed then quarks  and leptons 
are automatically  distinguished: The CKM matrix is unity: $V_{CKM} = 
V^{\dagger} V$,   while its leptonic analogue - the PMNS matrix \cite{pmns}  
$V_{PMNS} = V^{T} V$ -  contains two non-trivial mixing 
angles. These angles are related to  the structure of $V$ and can assume large 
values even when 
angles specifying the basic mixing matrix $V$  are relatively small, of the 
order of the Cabibbo angle. The corrections to basic ansatz are required 
in order to generate a non-trivial $V_{CKM}$.  The same  (or similar)  
corrections also modify $V_{PMNS}$, so that a kind of  complementarity 
relation appears in which quarks and lepton mixing angles  
sum up to  the angles of $V^{T} V$ instead to the maximal ones
as in \cite{qlc1}. 

We  discuss our ansatz  in the next section. 
Properties of the zero order  mixing are described subsequently in section 
3 and 4. Section 5 contains analysis of possible corrections  to these 
ansatz  assuming that  departures from universality in the quark and 
lepton sectors are 
correlated. We consider a specific scheme of perturbation in section 6 
which leads to such departures. 
A possibility of embedding  scenario into Grand Unification theory 
is outlined in section 7. Some comments on this  approach and summary are
contained in section 8.

\section{Ansatz}

Our ansatz is based on the relations among fermionic masses
that follow in grand unified theories. The minimal version of $SU(5)$
implies relation (\ref{su5}) and a symmetric up quark mass matrix $M_u$.
The Dirac neutrino mass matrix $M_{\nu D}$ gets related to $M_u$ in $SO(10)$
models - the minimal version giving
\be \label{so10} M_u^T=M_u=M_{\nu D}~ .\ee
It is possible to obtain $SO(10)$ models which 
simultaneously satisfy equalities (\ref{su5},\ref{so10}). 
We first consider implications of the above mass relations
for mixing and then make additional plausible assumptions to arrive at
a universal mixing ansatz.

We assume that to leading order in some approximation, the upper 
(up quarks, neutrinos) and the lower 
(d quarks, charged leptons) components 
of all the weak doublets are diagonalized by (complex) 
unitary matrices $V'$ and $V$ respectively. Specifically, assume
\beqa \label{ans1}
V'^{\dagger} M_u^0 V^{'*}=D_u^0&;& V'^{\dagger} M_{\nu D}^0 V'=D_{\nu D}^0 \nonumber ~,\\
V^{\dagger} M_d^0 V=D_d^0&;& V^T M_l^0 V^*=D_l^0~.  \eeqa
$D_f^0$ above refer to the diagonal mass matrices of fermions 
and superscript $0$ is used to denote the leading order. 
The first of these relations is based on the $SO(10)$  equality (\ref{so10})
which in other way around follows  from it if $D_u^0 = D_{\nu D}^0$. Likewise, 
relation (\ref{su5}) follows from eq. (\ref{ans1}) when $D_d^0=D_l^0$.
An assumption of a hermitian $M_d$ 
is made here which does not appear in $SU(5)$ but it can be achieved 
by imposing the left-right symmetry in addition. The light Majorana neutrino mass matrix is
assumed to be diagonalized in the same way as up-quarks:
\be
\label{mat-nu}
V^{'\dagger} M_{\nu}^0 V^{'*} = D_{\nu}^0.
\ee
The later can be reproduced by the seesaw mechanism 
if the Majorana mass matrix of the RH 
neutrinos  $M_R$ is also diagonalized as $M_{\nu}^0$: 
\beqa
\label{mat-ss}
V^{'\dagger } M_R^0 V^{'*}   &=& D_R^0 ~. 
\eeqa
Then for light neutrinos we have (in the basis $\bar{\nu} \nu^*$)
\be 
\label{mnu0}
M_{\nu}^0  = -M_D^0M_R^{0^{-1}}M_D^{0T} = -V' D_D^0 D_R^{0_{-1}} D_D^0 V^{'T} 
\ee
which satisfies eq.(\ref{mat-nu}).

Eqs.(\ref{ans1},\ref{mat-nu}) imply that in the leading order 
the mixing matrix for quarks equals 
\be
\label{phy0q2}
V^0_{CKM} = V'^{\dagger}V, 
\ee
and for leptons
\be
\label{phy0l2}
V^0_{PMNS} = V^{T} V' ~.
\ee
These equalities lead to the relation 
(apparently  $V' = V V^{0\dagger}_{CKM}$): 
\be
\label{re1}
V^0_{PMNS} = V^{T} V V^{0\dagger}_{CKM}. 
\ee
For complex $V$, $V^TV$ is non trivial with the result that
the lepton and quark mixings can be different in spite of unifying
relations among their mass matrices. 
From (\ref{re1}) we obtain 
\be
\label{rel1}
V^0_{PMNS} V^0_{CKM} = V^{T} V 
\ee
which shows a kind of  the quark-lepton complementarity: 
quark and lepton mixing angles add up to the 
angles of the  $V^{T} V$ matrix. This complementarity differs from the 
one considered in the literature \cite{qlc1,qlc2} in which $V^0_{PMNS}$ is assumed to have
the bi-maximal or tri-maximal \cite{trib} form and $V_{CKM}$ arises 
as a correction to it. It is not always possible to obtain
such mixing in a natural way while the relation (\ref{re1}) follows 
naturally from the GUT relations supplemented with additional assumptions
of hermitian $M_d$ and equality (\ref{mat-ss}).

The smallness of the CKM mixing angles  
requires $V'\sim V$. This suggests a universal ansatz in which $V'$ is 
equated to $V$ to leading order in some approximation.
This ansatz provides a universal description of all fermion mixings 
with the property that the leptonic mixing 
remains non-trivial,  $i.e.$, substituting $V'=V$ in eq. (\ref{phy0l2}), 
we obtain $V^0_{CKM} = V^{\dagger} V = I$ and 
\be 
\label{re2} V^0_{PMNS}=V^TV~. 
\ee

 Let us denote by $\tilde{f}$ the fermion basis which corresponds  
to the universal description. Then from eq. (\ref{ans1})  we obtain that  
$\tilde{f}$ are related to the mass basis as
\be \label{relation}
\tilde{d}_L=V d_L~~~;~~~
\tilde{l}_L=V^* l_L~~~;~~~
\tilde{f}_L=V f_L~,~~~({f=u,\nu} )~. \ee
 Let us emphasize that the different transformation properties of the charged lepton 
(compared to down quarks)  are  responsible for generating a non-trivial PMNS matrix. 
We shall designate this ansatz as version I. 
 
In an alternative version - to be called version II   
of the ansatz  the neutrino transformation property is different compared 
to the up-quarks. This version is based on the unified mass formula 
in the minimal $SO(10)$ according to which $M_l$ and $M_d$ are also 
symmetric and are proportional to the up quark mass matrix. 
So,  they are all diagonalized by the same matrix and thus $V=V'$:  
\be
\label{mat-f}
V^{\dagger} M_f^0 V^* = D_f^0, ~~~ f = u, ~d,~ l. 
\ee
Also the Dirac mass matrix of neutrinos may have the same property as 
in (\ref{mat-f}). It is however assumed that the left handed neutrino components 
transform differently compared to eq. (\ref{relation}): 
\be 
\label{neutrino}
\tilde{\nu}_L=V^*\nu_L, 
\ee
so that   eqs.(\ref{mat-f},\ref{neutrino}) now imply 
$ V^0_{CKM} = V^{\dagger}V =I,$ and 
%
\be
\label{phy01}
V^{0'}_{PMNS} = V^{\dagger} V^* ~. 
\ee
In this version the PMNS matrix 
differs from the earlier version (eq. (\ref{re2}))
by complex conjugation.

The  different transformation properties 
of the neutrino compared to the up-quark may arise from some special 
structure of $M_R$. However more appealing possibility is that the usual 
seesaw (type-I) just suppresses the neutrino Dirac mass 
and the main contribution to the neutrino mass matrix comes from the type-II
seesaw \cite{seesaw2}  with
\be \label{type2}
V^T M_\nu^0 V=D_\nu^0~. 
\ee

Let us underline several points in connection with  the proposed ansatz.
\begin{itemize}

\item The quark-lepton universality can be reconciled with difference of 
quark and lepton mixings. In contrast to previous 
proposals of the quark-lepton universality \cite{demo,sd} this difference appears already 
in the lowest order. 

\item 
The ansatz   (\ref{relation})  distinguishes  between  the charged leptons and other fermions if 
$V$ is 
complex. 
The GUT relation (\ref{su5}) provides  a basis  of introducing this distinction 
which results in 
different mixing patterns between quarks and leptons.

\item 
Equation (\ref{su5}) is known \cite{rev,lop1} to lead  to  
different patterns of quark and lepton mixing if $M_d^0$ has 
non-symmetric and lopsided form.  The lopsided  form is not required in the present ansatz 
and hermitian 
$M_d^0$ as implied by eq. (\ref{ans1})  is sufficient to produce the differences in 
quark and leptonic  mixings.

\item  In the ansatz II special form  of transformations of  neutrinos    
can be related to different origin (e.g. see-saw type II) and smallness of neutrino masses.

\item 
Mixing among fermions is universal but their mass matrices 
need not be so unless all $D_f^0$
are proportional  and $V$ is real. In view of  the strong hierarchy 
in masses, it is natural 
to assume  that
only the third generation is massive in which case, $M_f^0$ for all $f$
can be related to a universal matrix as we will show. 

\item 
The ansatz remains unspecified until $V$ is given. 
If $V$ were to describe only quark mixing then different  choices 
of $V$ would be equivalent at the leading order. 
But due to eqs. (\ref{relation},\ref{neutrino}), $V$ acquires  
a physical content and gets related to $V_{PMNS}$. This 
allows direct experimental determination of the universal basis 
unlike other universality ansatzs \cite{demo,sd} proposed before. 

\end{itemize}

Since $V_{CKM}=I$ to the lowest order, some additional sources of fermion masses should 
exist. 
In general, they will also contribute to lepton mixing. 
We assume that this additional mixing is small - of the order of the Cabibbo mixing, so that 
$V^0_{PMNS}=V^TV$ can be considered as the lepton mixing in the lowest order approximation.

\section{Properties of the zero order mixing matrix }

Let us explore properties of the mixing matrix $V_{PMNS}^0 \equiv V^T V$. 
The results are applicable to both versions of the universality. 
Apparently $V_{PMNS}^0$  is a symmetric unitary matrix 
which can be parameterized\footnote{The above
parameterization was used in \cite{branco} to describe mass matrix
for the degenerate neutrinos. Here it describes mixing matrix. Our
choice differs from the one in \cite{branco} by  an interchange
$\th\rightarrow -\th$.} as 

\beqa 
\label{mns2} 
V^0_{PMNS}&=& P~U~P \nonumber \\
U&=&\left(\ba{ccc} c_\th&-c_\phi s_\th&-s_\phi s_\th \\ -c_\phi
s_\th&e^{i \a} s_\phi^2-c_\phi^2 c_\th&-c_\phi
s_\phi(c_\th+e^{i\a})\\ -s_\phi s_\th&-c_\phi
s_\phi(c_\th+e^{i\a})&e^{i\a} c_\phi^2-c_\th s_\phi^2\\ \ea
\right)  ~,\eeqa
where $c_\th \equiv \cos \theta$, $c_\phi \equiv \cos \phi$,  etc., 
$\theta$,  $\phi$ and $\alpha$ 
are new free parameters.  $P$ is a phase matrix:
\be 
\label{p} 
P={\rm diagonal} (e^{i \rho},e^{i \sigma},1)~ .
\ee
The matrix $P$ on the left hand side of $U$ in eq. (\ref{mns2}) can be rotated away 
by redefining  the charged lepton fields. The one on the right 
hand side contains the Majorana phases associated with the neutrino fields.

The $V^0_{PMNS}$  depends on two instead of three mixing 
angles which implies
restrictions on the physical parameters. Comparing eq.(\ref{mns2}) with 
mixing matrix in the standard parameterization \cite{rev} we obtain the 
following expressions for the standard mixing angles: 
\beqa 
\label{qlcrel} 
\tan \theta_{12} & = & - \cos\phi \tan \theta , 
\nonumber \\
\sin \theta_{13} & = & - \sin \phi \sin \theta , 
\nonumber \\
\sin \theta_{23} & = & - \frac{\sin \phi \cos\phi (\cos \theta + 
e^{i\alpha})}{\sqrt{1 - \sin^2\phi \sin^2\theta}} . 
\eeqa
\begin{figure}[h]
\centerline{\psfig{figure=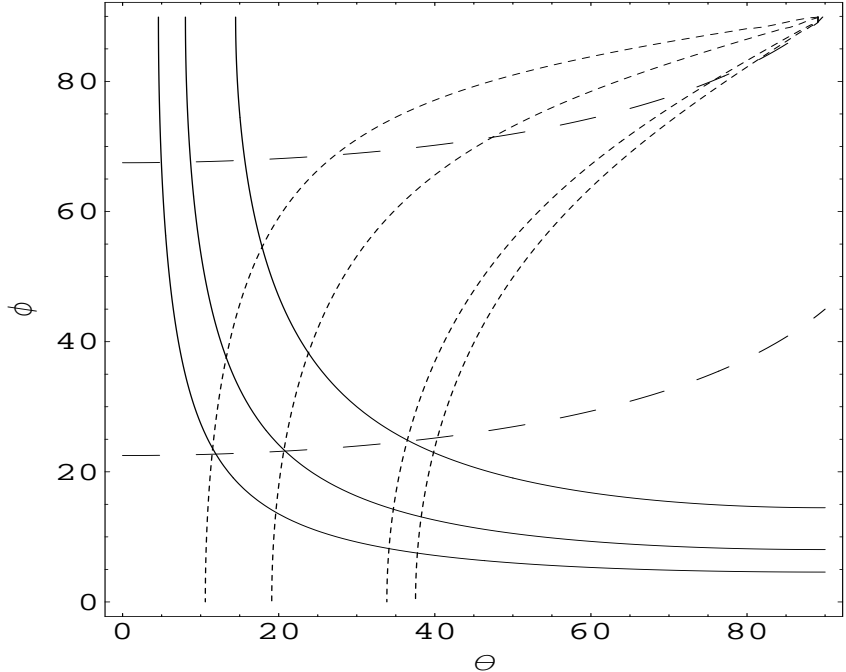,height=10cm,width=12cm,angle=0}}
\vskip  0.5cm
\caption{The contours of constant values of neutrino oscillation parameters  
in the $\phi-\th$ plane  for a symmetric PMNS matrix. 
The dotted curves correspond  to $\theta_{12}=10^\circ({\rm left~ 
most}),19^\circ,34^\circ,37.5^\circ$.
The dashed curves correspond  to $\sin^2 2 \theta_{23}=1.0$ 
and the solid  ones - to $\sin\theta_{13}=0.08$ (left~ most), $0.14,~ 0.25$.}
\end{figure}
\vskip 0.5cm

Using these relations we find lines of constant values of 
the observables in the $\phi - \theta$ plane which are shown in Fig. 1. 
Notice that only $\theta_{23}$ depends on the phase  $\alpha$. 
The~~ Fig. 1 corresponds to $\alpha = 0$. With increase in
$\alpha$ the angle $\theta_{23}$ decreases and therefore the corresponding 
iso-contours shift to larger  $\phi$ and  $\theta$.

An important consequence of the
restricted form for $V_{PMNS}^0$ is that it can not 
reproduce the  bi-maximal (as well 
as tri-bimaximal \cite{trib}) mixing matrix which implies zero 1-3 mixing. 
Indeed, according to  
(\ref{qlcrel}), the condition  $\sin \theta_{13} = 0$ 
requires  that either the solar  or the atmospheric angle  vanishes. 
Conversely, $\theta_{12} = \theta_{23} = 45^{\circ}$ 
lead to unacceptable value $\sin \theta_{13} \approx 0.7$  
unless some other sources of masses 
and mixing compensate this zero order mixing. 

$V_{PMNS}^0$ can, however,  provide a good zeroth order approximation.
In particular, it can be a source of single maximal mixing. 
For maximal 2-3 mixing  
we find the isoline $\tan \theta_{23} = -1$:  
\be
\cos \theta = \frac{1 - \tan \phi}{\tan\phi(1 + \tan\phi)}. 
\ee
The central measured values $\theta_{12} = 34^{\circ}$ and 
$\theta_{23}= 45^{\circ}$ can be achieved at 
$\phi = 25^{\circ}$ and $\theta = 37^{\circ}$. This gives 
$\sin\theta_{13} =  0.25$ which is above the $3\sigma$ allowed region. 
However corrections to the 1 - 3 mixing of the Cabibbo angle size, 
$\Delta \theta_{13} \sim \theta_C = 13^{\circ}$ 
can easily  lead to the acceptable values. 
The best fit isolines for $\theta_{12}$ 
and $\theta_{23}$ also cross in Fig. 1 at 
$\phi=76^\circ,~~\theta =80^{\circ}$, 
which implies  $\sin\theta_{13} = 0.95$. 
 The latter is too large to be properly 
corrected by the Cabibbo angle scale.  

According to Fig. 1, the $1\sigma$ acceptable values of  1-3 mixing,  
$\sin\theta_{13} < 0.14$,  
require (for maximal 2-3 mixing) that $\theta_{12} \leq  19^{\circ}$.  
So,  additional contribution to the 1-2 mixing  
of the order $\theta_C$ from some other source can bring the 1-2  mixing angle 
back to the central measured  value. 
Notice that for such a possibility $\theta \sim 21^\circ,~\phi \sim 23^{\circ}$. 
Next generation of the experiments (Double CHOOZ \cite{dchooz}, JPARC\cite{jparc}) 
may improve the 
bound on 1-3 mixing down to $\sin\theta_{13} < 0.08$ ($90 \%$ C.L.).  In this case 
(see Fig.  1) the maximal   2-3 mixing would require $\theta_{12} \leq 
10^{\circ}$ which is not possible to correct by the Cabibbo-type 
rotation. 

For non-zero $\theta_{12}$ and  $\theta_{23}$ 
the lower bound on $\sin \theta_{13}$ appears which 
can be quantified in the following way. 
According to 
eqs. (\ref{mns2}, \ref{qlcrel}) 
\beqa 
\label{inequality} 
|\sin \theta_{13}|&\geq& \frac{|V_{\mu 3}|}{|V_{e2}|} (1-|V_{e1}|)
\nonumber \\ &\geq&  \frac{|\sin\theta_{23}|}{|\sin\theta_{12}|}
(1-|V_{e1}|) \geq \frac{|\sin\theta_{23}|}{|\sin\theta_{12}|} 
(1-|\cos \theta_{12}|),
\eeqa
where equality holds if $\alpha=0$. Here $V_{\alpha i}$ are the elements of the 
PMNS matrix.  For central values 
of the mixing angles, eq. (\ref{inequality}) gives $|\sin \theta_{13}| \geq 0.23$. 
This bound weakens with decrease of $\theta_{12}$ and $\theta_{23}$. 
Taking the $3\sigma$ lower values for these angles we find from 
eq. (\ref{inequality}): $|\sin \theta_{13}| \geq 0.16$ which is at the $2\sigma$ 
upper bound. 

For non-zero $\alpha$ the 1-3 mixing equals 
\be
\label{alpha} 
|\sin \theta_{13}| = 
\frac{|V_{\mu 3}|}{|V_{e2}|} (1-|V_{e1}) 
\left[ 1 - \frac{2\cos\theta (1 - \cos\alpha)}{(1 + 
\cos\theta)^2}\right]^{-1/2} . 
\ee
One can see  that  expected value of 
$\sin \theta_{13}$ increases  with $\alpha$,  thus worsening the overall fit. 

The matrix $V_{PMNS}^0$ alone (without corrections) does not 
reproduce well the observed lepton mixing matrix. 
The best global fit of the data would correspond to 
$(2 - 3) \sigma$ lower values of $\theta_{12}$ and  $\theta_{23}$  
and $\theta_{13}$  at $(2 - 3) \sigma$ upper bound. 
This fit corresponds to 
\be 
\label{range1} 
\phi \approx 19^{\circ} - 25^{\circ}~;~~~~\th \approx 30^{\circ} - 40^{\circ},  
~~~~\alpha = 0, 
\ee 
which can be read-off from Fig. 1. Clearly region above 
the upper bound on 1 -3 mixing worsens the fit. 

The neutrino masses that follow from the leading order expression, 
eq.(\ref{mnu0})
are immediately determined by the eigenvalues of the Dirac and  
Majorana mass matrices:  $D^0_{\nu} = - D^0_{D}D^{0 -1}_{R}D^0_{D}$.  In 
particular, for the ratio we obtain:
\be 
\label{hier1}
\frac{m_{\nu_2}^0}{m_{\nu_3}^0} = 
\left(\frac{m_{2D}^0}{m_{3D}^0}\right)^2~\frac{M_3^0}{M_2^0} ~.
\ee

\section{Universal mixing and universal mass matrices. }

The phenomenological determination of $V^0_{PMNS}$ 
can be directly used to  obtain $V$ and hence the 
fermion mass matrices in the lowest order in the universal basis. By definition, $V$  
satisfies the equality 
\be 
\label{v1} 
V^*V^0_{PMNS}V^{\dagger}= I  
\ee
which means that $V$ can be determined from diagonalization of  $V^0_{PMNS}$.  
In this way we find 
\be 
\label{v} 
V=D^*\tilde{V} P ,  
\ee
where $P$ is given by eq. (\ref{p}); 
the diagonal  matrix $D$ 
\be
D = diag(1,i,1) 
\ee
plays crucial role in generation of the non-zero lepton mixing and 
\be 
\label{tildev} 
\tilde{V}=
\left(\ba{ccc} \cos \frac{\th}{2}&-\cos\phi \sin \frac{\th}{2} &-\sin\phi 
\sin \frac{\th}{2} \\ 
- \sin \frac{\th}{2}&- \cos\phi \cos \frac{\th}{2}& - \sin\phi \cos 
\frac{\th}{2}\\ 
0&\sin\phi e^{i\a/2}&-\cos\phi e^{i \a/2}\\ \ea
\right).  
\ee
It can be represented as $\tilde{V} = Z_{\alpha} R_{12}(\theta/2) 
R_{23}(\phi) Z_1$, where $Z_{\alpha} \equiv diag(1, 1, e^{i\a/2})$, 
$Z_2 \equiv diag(1, - 1, -1)$. 
The matrix $V$ is determined from $V^0_{PMNS}$ only up to an orthogonal 
transformations $O$,  
since $V$ and $OV$ both lead to the same $V^0_{PMNS}$ 
and unit $V_{CKM}$. All leading 
structures differing by $O$ are thus physically equivalent 
and we shall choose $O=1$.

Notice that the basic mixing angles appearing in  eq.(\ref{tildev}),   
$\phi \sim \theta/2 \sim {\cal O}(15^\circ-20^\circ)$,  
are  relatively small - 
not far away from the Cabibbo  angle. 
Starting with such a small mixing, the above ansatz  is able to reproduce 
the leptonic phenomenology reasonably well.

Using relations  (\ref{ans1} - \ref{mat-ss}) (with $V'=V$) 
and expression for  $V$ (\ref{v}) we  can 
write the mass matrices in the  universal basis as 
\be
\label{matr-u}
M^0_{up} =  D^*\tilde{V} P^2 D_{up}^0\tilde{V}^T D^*, ~~~~ (up = u, \nu D,R) , 
\ee
\be
\label{matr-dl}
M^0_d = D^*\tilde{V} D_{d}^0\tilde{V}^\dagger D, ~~~~~~
M^0_l = D\tilde{V}^* D_{l}^0\tilde{V}^T D^*, 
\ee
for the version I  and
\be
\label{matr-f}
M^0_f = D^*\tilde{V} P^2 D_{f}^0\tilde{V}^T D^*, ~~~~ (f = u, d, l), 
\ee
\be
\label{matr-nu}
M^0_{\nu} =  D \tilde{V}^*P^{*2} D_{\nu}^0\tilde{V}^\dagger D, 
\ee
for the version II. 

Natural origin  of the universality of mixing 
proposed in this paper would be  
universality of the mass matrices, at least in some approximation. 

According to (\ref{matr-u} - \ref{matr-nu}), $M_{f}^0$ for all 
fermions can be related to the same universal mass matrix in the 
approximation in which only the third generation fermions are massive, {\it i.e.} 
$D_f^0 = diag(0, 0, m_{3f})$ and $\alpha = 0$.
(The later is required by phenomenology.)
In this case, eqs. (\ref{ans1},\ref{mat-ss}) and (\ref{v}) give
%
%
\be
\label{matr-univ1}
M^0_{u} = m_{3u} D^*\tilde{M}D^*,  ~~
M^0_{\nu} = m_{3\nu} D^*\tilde{M} D^* , ~~
M^0_d = m_{3d}  D^*\tilde{M} D, ~~
M^0_l = m_{3l} D \tilde{M} D^*,
\ee
in the first version,  and 
\be
\label{matr-univ2}
M^0_{f} = m_{3f} D^*\tilde{M}D^*, ~~(f = u, d, l), ~~~~~~ 
M^0_{\nu} = m_{3\nu} D \tilde{M} D 
\ee
in the second version. 
The universal matrix  $\tilde{M}$   can be written as
\be \label{mtilda1}\tilde{M}=\left( \ba{ccc}
s_\phi^2 s_{\frac{\th}{2}}^2 & s_\phi^2s_{\frac{\th}{2}} c_{\frac{\th}{2}}&s_\phi c_\phi 
s_{\frac{\th}{2}} \\
s_\phi^2 s_{\frac{\th}{2}} c_{\frac{\th}{2}} & s_\phi^2c_{\frac{\th}{2}}^2 &s_\phi c_\phi 
c_{\frac{\th}{2}} \\
s_\phi c_\phi s_{\frac{\th}{2}} & s_\phi c_\phi c_{\frac{\th}{2}}&c_\phi^2 \\
\ea \right)~, 
\ee
$s_{\frac{\th}{2}} \equiv \sin 
\frac{\th}{2}, ~~c_{\frac{\th}{2}} \equiv \cos\frac{\th}{2}$.

All fermionic mass matrices are determined 
by a universal matrix  $\tilde{ M}$ as in 
the previous ansatz \cite{demo,sd}. But unlike them, the presence  of $D$ in 
eqs. (\ref{matr-univ1},\ref{matr-univ2}) changes the mixing patterns and allows 
large mixing for leptons even before perturbations are introduced. 

If only partial universality 
is assumed: $V'\not  =V$,  
then from eq. (\ref{ans1}) we find
$(M_{up}^0)_{ij}=m_{3up}V'_{i3}V'_{j3}$. One can choose a parameterization for $V'$ in such a way that 
$M_{up}^0$ gets related to a matrix having the same form
as in eq. (\ref{mtilda1}) but with different mixing angles.

We can further absorb the complex matrices $D$
by the redefinition of fields: 
\be
\tilde{f} = D^* f' ~~(f = u_L, u^c_L, d_L, d^c_L, \nu_L), ~~~ 
\tilde{l}_{L,R} = D {l'}_{L,R}. 
\ee
In the basis $f'$, $l'$, all the matrices equal $\tilde{M}$. 
Notice that in the new basis the charge current for the 
second generation of leptons has an opposite sign with respect to 
all other currents.

Since the universal matrices are singular (rank 1)
this scenario can not be realized in the 
context of the see-saw
with the same universal structure of $M_R$. One needs to assume that 
$M_R$ is non-singular but is  still diagonalized by the same 
rotation as $M_{\nu D}$. 

 Let us consider the structure of the universal matrix $\tilde{M}$. 
Define
$\e_1\equiv \sin\phi$ and $\e_2\equiv \sin\frac{\th}{2}$.
According to   eq.(\ref{range1})  both $\e_1$ and $\e_2$ are 
relatively small  parameters,
$0.3\leq\e_1\leq 0.4$ and $0.25\leq \e_2\leq 
0.34$. Then  $\tilde{M}$ assumes the following form if only the leading 
terms in $\e_1,\e_2$ are kept:
\be \label{mtilda2}\tilde{M}=\left( \ba{ccc}
\e_1^2 \e_2^2&\e_1^2 \e_2&\e_1\e_2 \\
\e_1^2\e_2&\e_1^2 &\e_1 \\
\e_1\e_2&\e_1&1 \\
\ea \right).  \ee

An ansatz very similar to eq. (\ref{mtilda2}) was proposed in \cite{sd}
for all fermion masses. 
The $\tilde{M}$  in eq. (\ref{mtilda2}) coincides with the 
ansatz in \cite{sd} if one identifies  $\e_1$ and  $\e_2$ with the parameter 
$\lambda$ of \cite{sd}. The phenomenologically
required values for $\e_1,\e_2$ are not 
very different from the value $\lambda\sim 0.26$ used in \cite{sd}. 
In the present case, the above form for  $M_f^0$ arises from the 
basic  ansatz and $V^0_{PMNS}$ determines $\e_1,\e_2$. 

The specific form in eq. (\ref{mtilda2}) has interesting consequences 
which were elaborated in \cite{sd}.  
One naturally gets hierarchical masses and the 
correct hierarchy in the quark mixing  
angles when small perturbations are added to 
this ansatz. We shall come back to these in sect. 6.

Another possibility to introduce the universal mass matrix is to 
assume that in the lowest approximation $P^2 = I$, and 
for all fermions the diagonal mass matrices have the same 
hierarchy of the eigenvalues: 
$D_0 \equiv diag(m_1, m_2, m_3)$.  Then the universal matrix 
equals 
$\tilde{M} = \tilde V D_0 \tilde V^T$. 
In this case the RH neutrino mass matrix can also have the 
universal form.

Two different universal diagonal matrices for up and down 
particles $D^0_{up}$ and $D^0_{down}$     
can be introduced in the context of partial universality, eq. (\ref{ans1}).

\section{Corrections to  $M_f^0$ and mixing}


The above ansatz needs corrections to  account  for the non-trivial quark 
mixing. Similar corrections 
to the leptonic sector would change the PMNS matrix  as well. These two 
corrections may be independent  or may also show some quark-lepton symmetry 
thus leading to kind of complementarity related to the 
zero order mixing $V^0_{PMNS}$. 
Notice that in  \cite{qlc1} a bi-maximal 
form  for $V^0_{PMNS}$ was assumed which leads to
numerical relations between the solar and the 
Cabibbo angle  such as $\theta_\odot=\pi/4-\theta_C$. 
In the present case, the quark and lepton mixings 
differ by the angles of the zero order matrix.

Let us add a perturbation $\delta M_f$ to $M_f^0$ and consider 
\be  M_f=M_f^0+\delta M_f ~.\ee
$\delta M_f$ would lead to additional mixing $V_f$. 
Referring to the first scenario (sect. 2)  we define,
\beqa 
\label{fullf} 
V_f^{\dagger} V^{\dagger} M_f V^* V_f^* &=&D_f~~~~~ f=u,\nu D,R ~,\nonumber\\
V_d^{\dagger} V^{\dagger} M_d V V_d&=&D_d~,\nonumber\\
V_l^{T} V^{T} M_l V^* V_l^*&=&D_l ~, 
\eeqa
where $D_f$  are the diagonal mass matrices after corrections.
For simplicity, we have assumed $\delta M_f$ to have the same symmetry properties 
as $M_f^0$,  $i.e.$, symmetric for $f=u,\nu D,R$ and hermitian for $f=d,l$. 
Eq. (\ref{mnu0}) now gets replaced by
\beqa \label{mnu} {\cal M}_\nu&=&-M_{\nu D}~M_R^{-1}~M_{\nu D}^T ~, \nonumber \\
&=&- V~V_{\nu D} {\cal F} D_\nu {\cal F}^T V_{\nu D}^T V^T ~, \eeqa
where ${\cal F}$ satisfies
\be \label{scriptf} {\cal F} D_\nu  {\cal F}^T=D_{\nu D} (V_{\nu D}^T V_R^*) 
D_R^{-1} 
(V_{\nu D}^{T} V_R^*)^T D_{\nu D} 
\ee
and $D_\nu$ is a diagonal matrix of the light neutrino masses.
It follows from the above that 
$$ V_\nu=V V_{\nu D} {\cal F}. $$
This together with eq. (\ref{fullf}) give us
\beqa \label{phy} V_{CKM}&=&V_u^{\dagger}V_d ~, 
\nonumber\\ V_{PMNS}&=&V_l^{T} V^0_{PMNS}V_{\nu D}{\cal F} ~. 
\eeqa
The $V_{CKM}$ and corrections to $V^0_{PMNS}$ depend upon the 
perturbations. 
One can assume  that they have similar origin and 
turn out to be  of the same order. 

Here we explore phenomenological consequences 
of  assumption  that the leading corrections to mixing display 
a quark-lepton symmetry.  

Consider, {\it e.g.}, the following simplest possibility:  

\noindent ($i$) The CKM matrix arises essentially 
from corrections to the down quark matrix: $V_d\approx V_{CKM}$ and 
$V_u\approx 1$.\\
\noindent ($ii$) 
The charged lepton and the Dirac neutrino mixing matrices 
satisfy $ V_d \approx V_l^*$ (which corresponds to the zero order relation)  and 
$V_{\nu D}=V_u\approx 1$.\\   
\noindent ($iii$) $M_R$ receives very small 
corrections, so that $V_R\sim 1$.  \\[.25cm]
\noindent The above assumptions imply ${\cal F}\approx 1$ 
in eq. (\ref{scriptf}) and the following relation 
\be 
V_{PMNS}=V^{\dagger}_{CKM} V^0_{PMNS} ~.
\label{pmns1}
\ee
The $V_{PMNS}$ is no longer a symmetric matrix because of the
presence of the CKM matrix on the left. Also, the matrix $P$
on the left of eq.(\ref{mns2}) now cannot be absorbed directly in redefining
the charged lepton fields and can be physically relevant. 
The results presented below assume $P=1$.  

The eq. (\ref{pmns1}) can be rewritten as 
$V_{CKM} V_{PMNS} = V^0_{PMNS}$ showing a kind of 
complementarity of the quark and lepton mixings. 

In the first approximation we can neglect 2-3 and 1-3 quark  mixing 
considering $V_{CKM} \approx V_{12}(\theta_C)$. 
We then obtain: 
\be
V_{PMNS} = V^{\dagger}_{12}(\theta_C) U (\theta, \phi) ~, 
\label{pmnscor}
\ee
where $U(\theta, \phi)$ is given in (\ref{mns2}). 
The additional 1-2 rotation in (\ref{pmnscor}) modifies all zero order 
mixing angles. It is easy to show  that for 
large (nearly maximal) 2-3 mixing, this rotation simultaneously 
increases (or diminishes) $\theta_{12}^0$ and $\theta_{13}^0$. 
Due to smaller value $\theta_{13}^0$, the relative changes of the 
1-3 mixing is stronger than 1-2 mixing. 
Therefore by $\sim \theta_C$ rotation one can improve fit of the data 
in comparison with zero order mixing by

1) taking larger values of $\phi$ and $\theta$ than in (\ref{range1}); 
this will lead to larger zero order values of $\theta_{12}^0$ and 
$\theta_{13}^0$;  

2) performing  $\theta_C$ rotation which reduces 1-2 and 1-3 mixing. 
Since reduction of the 1-3 mixing is stronger, we  obtain smaller 
than in the previous case  values 
of $\theta_{13}$ for the allowed values of $\theta_{12}$.   
\begin{figure}[h]
\centerline{\psfig{figure=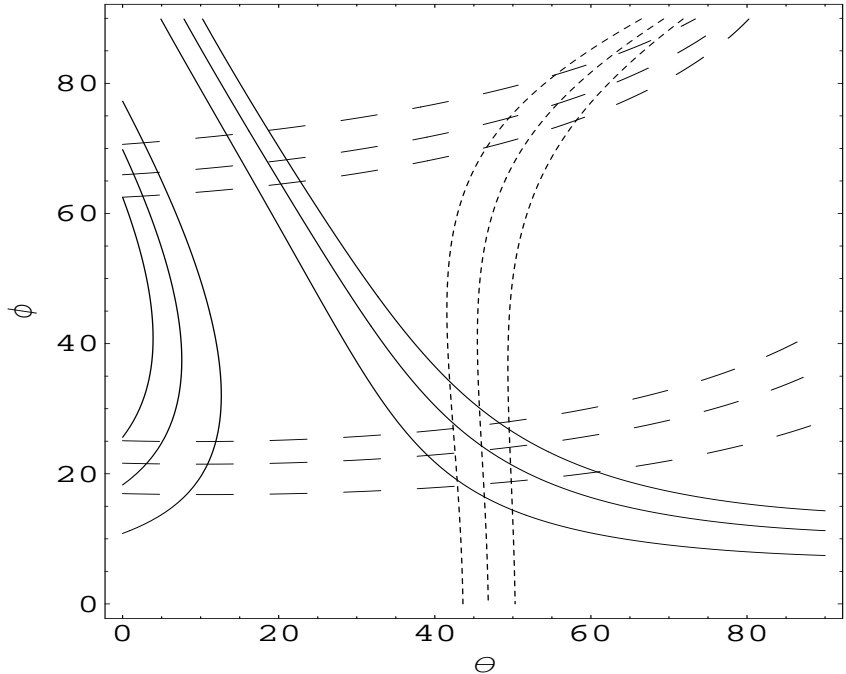,height=10cm,width=12cm,angle=0}}
\vskip  0.5cm
\caption{The contours of constant values of  
the neutrino oscillation parameters  
in the $\phi-\th$ plane 
in the presence of the perturbations  
specified in the text. 
The dotted curves are for $\theta_{12}=30.6^\circ,34^\circ,37^\circ$.
The dashed curves correspond  to $\sin^2 \theta_{23}=0.34,~0.5,~0.62$ 
and the  solid curves to $\sin\theta_{13}=0.09,~0.14,~0.17$.
 Central lines correspond to the best fit values and the outer ones restrict  
the 2$\sigma$ allowed regions.  }
\end{figure}
\vskip 0.5cm
These features can be seen in Fig. 2,  
where we show $2\sigma$ 
allowed ranges of observables in the $\phi-\th$ plane.
The full $V_{CKM}$ instead of $V_{12}(\theta_C)$ is used in drawing this figure. 
Unlike in the zeroth order case, 
now a sizable intersection region occurs 
at 2$\sigma$ which  corresponds to
\beqa 
\label{90range}
\phi \approx 20^\circ-28^\circ, & ~&~~~~\th\approx 42^\circ-48^\circ ~, \nonumber \\
|\sin\theta_{13}|&\sim& 0.09-0.17~. 
\eeqa
We find that at 3$\sigma$ level  $\sin\theta_{13}$ still has a lower bound $|\sin\theta_{13}|\sim 0.08$  
which can be tested in the next generation of experiments. 
The best fit values of observables $\tan^2\theta_{12}= 0.45,~\sin^2 2 \theta_{23}=1$ 
and  $\sin\theta_{13}=0.13$  are obtained for $\th = 46.01^{\circ}, \phi=23.6^\circ$. 

In case of the partial universality, $V_{PMNS}$ is obtained by the right multiplication
of $V^T V$ with $V_{CKM}^\dagger$, eq.(\ref{re1}). 
In this case, starting with initially negative $\tan \theta_{12}$, 
the $V_{12}^\dagger(\theta_C)$ rotation increases 
the value of $\tan^2\theta_{12}$ without affecting $\theta_{13}$. As a consequence, one 
can obtain a good fit to data by initially choosing smaller 
values of $\phi,\theta$ corresponding to a smaller $\theta_{13}$ and $\theta_{12}$. 
Cabibbo rotation then increases the solar angle to the required value. Indeed, one finds
(again using full $V_{CKM}$) that eq. (\ref{re1}) can reproduce 
the best fit values $\tan^2\theta_{12}=0.45,\sin^2\theta_{23}=0.5$, 
and predicts $\sin \theta_{13}=0.13$ 
with $\theta=22.35^\circ,\phi=24.34^\circ$.

\section{Universal mixing and mass hierarchies}

We now discuss specific perturbations to $M_f^0$ which approximately 
reproduce  the scenario outlined above. Following \cite{sd}, we assume that 
$(M_f^0)_{ij}$ receives  sub-leading corrections parameterized 
by $\delta_{ij}^f\leq {\cal O}(\e_1,\e_2)$, 
and the $\tilde{M}$ 
in eq.(\ref{mtilda2}) changes after perturbations to
\be \label{mtilda3}\tilde{M}_f=\left( \ba{ccc}
\e_1^2 \e_2^2(1+\delta_{11}^f)&\e_1^2 \e_2(1+\delta_{12}^f)&\e_1\e_2(1+\delta_{13}^f) \\
\e_1^2\e_2(1+\delta_{12}^f)&\e_1^2 (1+\delta_{22}^f)&\e_1(1+\delta_{23}^f) \\
\e_1\e_2(1+\delta_{13}^f)&\e_1 (1+\delta_{23}^f)&1 \\
\ea \right) ~.\ee
Here $f$ refers to all fermions and $\delta^f_{ij}$ are taken to be real.
These perturbations generate ($i$) masses for the  first two generations, 
($ii$) non-trivial CKM matrix,  and ($iii$) corrections to  $V_{PMNS}^0$.
The ratios of masses are given by \cite{sd}
\beqa \label{masses} 
\frac{m_2^f}{m_3^f}&\approx& \e_1^2(\delta_{22}^f-2 \delta_{23}^f)~, \nonumber \\
\frac{m_1^f}{m_3^f}&\approx& \e_1^2 \e_2^2\left(\delta_{11}^f-2 \delta_{13}^f-\frac{(\delta_{12}^f-\delta_{23}^f-\delta_{13}^f)^2}{\delta^f_{22}-2 \delta_{23}^f}\right)~.
\eeqa
The above relations hold for all fermions including the right handed neutrinos.

The corrected $M_f$   is diagonalized as in eq.(\ref{fullf}). The correction
$V_f$ to the zeroth order matrix can be found by approximately diagonalizing
$\tilde{M}_f$ through successive rotations \cite{sd}. This procedure leads to
\be \label{vf}
V_f\approx P^* \left( \ba{ccc}
1&-\e_2(1-h^f)&-\e_1\e_2(\delta_{23}^f-\delta_{13}^f)\\
-\e_2(1-h^f)&-1&-\e_1\delta_{23}^f\\
-\e_1\e_2(\delta_{23}^f h^f-\delta_{13}^f)&\e_1 \delta_{23}^f&-1\\ \ea \right)~, \ee
where $$h^f=\frac{\delta_{12}^f-\delta_{13}^f-\delta_{23}^f}{\delta_{22}^f-2 \delta_{23}^f}$$

The mixing matrices follow from the  eqs. (\ref{vf})  and (\ref{phy}).
As noted in \cite{sd}, the above $V_{d,u}$ automatically lead to hierarchical 
mixing angles for quarks, 
$V_{us}\sim \e,V_{cb} \sim \e \delta, V_{ub}\sim \e^2 \delta$. 
As regards the leptonic sector,  the basic mixing is still given by $\theta,\phi$ and 
perturbations generate additional corrections.

The main problem is to reconcile the proposed universality of mixing and 
strongly different mass hierarchies of quarks and leptons. 
Let us show that it is possible for particular choice of  
parameters $\delta$. 
 We assume that perturbations to $M_u^0,M_R^0,M_D^0$ satisfy
\be 
\label{relations} 
\delta_{12}^f=\delta_{22}^f~ \equiv \delta^f;~~~ \delta_{23}^f \approx \delta_{13}^f 
\approx 
0 ~.
\ee
In this case 
${m_2^f}/{m_3^f} \approx  \e_1^2\delta^f$ and $h^f = 1$.  
Consequently,  effect of corrections is only to induce masses for the first two 
generations as in eq. (\ref{masses}). 
The mixing remains unchanged to leading order and is given by $V$ since $V_f$ 
 (eq.~({\ref{vf}))
has only diagonal elements when conditions (\ref{relations}) are imposed. 
Due to  equality (\ref{relations}), correction to the 1-2 mixing 
appears in the higher order in $\epsilon$: 
\be
\Delta \theta_{12} \approx \epsilon_2^3 \left(\frac{\delta_{11}^f}{\delta^f} -1 
\right).  
\ee
If $\delta_{11}^f = \delta^f$ the correction is zero, but since $\delta^f \neq 0$ 
the mass of second eigenvalue can be changed arbitrary (within the application limits). 
This illustrates that certain symmetry (equalities) of corrections can lead 
to unchanged mixing but different hierarchies.

Perturbations to the down quark 
and the charged lepton mass matrices should generate the CKM mixing and therefore 
$\delta_{11,22}^f$ (for $f=d,l$) do not satisfy eq. (\ref{relations}). 
Freedom in these parameters is enough 
to reproduce the $V_{CKM}$, $V_{PMNS}$ and the mass hierarchies correctly.
We have checked numerically that this can be done with corrections $\delta^f \leq 
0.3$.  

\section{Universality and Unification}

Our ansatz is based on the eqs. (\ref{su5},\ref{ans1},\ref{mat-ss}). 
We now show that these relations can be obtained in an  $SO(10)$ model with some
assumptions.
Specific model that we discuss gives eq.(\ref{su5}) and the following
relations:
\begin{equation}
\label{so10relation}
M_u^0=M_u^{0T}=-1/3 M_{\nu D}^0=\beta  M_R^0 ~, \end{equation}
where $\beta$ is the ratio of the weak scale to the right-handed neutrino 
mass scale.
These equalities allow a single unitary matrix $V'$ to simultaneously diagonalize
$M_u^0,M_{\nu D}^0$ and $M_R^0$ as in eqs.(\ref{ans1},\ref{mat-ss}) and lead to
eq. (\ref{rel1}).
 
The eqs. (\ref{su5},\ref{so10relation}) arise from 
the following $SO(10)$ invariant couplings in the super potential:
\begin{equation} \label{w1}
 W_d=\frac{f_{ij}}{M} (16_i^T  H)(H'^T  16_j) +
\frac{f'_{ij}}{M} (16_i^T  H')(H^T 16_j)\end{equation}
\begin{equation}
\label{w2}
W_u=g_{ij}
16_i^T  16_j\Phi~.
\end{equation}
Here, $H,H'$ are 16-plets of Higgs; $16$ refer to the fermions and $i,j$ are generation indices. $\Phi$ is a 
$\overline{126}$-plet.
The multiplets in the parenthesis in eq.(\ref{w1}) are contracted to form
$SO(10)$ vectors.
Two remarks are in order.

\begin{itemize}
\item The superpotential (\ref{w1}) was used in \cite{lop1} to obtain 
eq. (\ref{su5}). Since the 
16-plet contains only the Higgs doublet $H_d$ with hypercharge $-1/2$, eq.(\ref{w1})
gives mass only to the charged leptons and the down quarks. 
The $H,H'$  contain neutral fields with the quantum numbers of $\nu,\nu^c$.
Vacuum expectation value (vev) for $\nu^c$ breaks $SO(10)$ preserving SU(5). This leads to 
$SU(5)$ relation eq. (\ref{su5}) when  $\nu,\nu^c$ components of $H,H'$ 
acquire vevs. 

\item $\overline{126}$ contains both the up type 
and the down type doublets, $ H_{u,d}$,  and triplets, $\Delta_{L,R}$ under 
respective $SU(2)$ groups.  Vacuum expectation value for 
$H_u$ field generates eq. (\ref{so10relation}). One needs to assume here that 
$H_d$ component in $\overline{126}$ has zero vev 
at the minimum or that its contribution 
to $M_d,M_l$ is sub-dominant compared to the contribution from the 
superpotential (\ref{w1}). 

\end{itemize}
This example demonstrates that  differences in quark and lepton mixings can naturally arise 
in $SO(10)$ scheme. The mixing matrices  satisfy 
the complementarity relation eq. (\ref{rel1}) if $M_d$ is Hermitian. It is 
non-trivial to derive
the relation $V'=V$ used in the ansatz and may require 
a larger quark-lepton symmetry not envisaged in $SO(10)$.  


\section{Discussion }

We discuss here some open questions.
The proposed ansatz requires complete universality and treats the right 
handed neutrinos also on similar footings as other fermions in version I. 
Actually the right handed neutrino masses may have different origin and may not show 
complete universality.
Some of the above considerations remain valid even if
the right handed neutrinos display different mixing pattern, $i.e.$, 
\be \tilde{\nu}_R=\tilde{V}_R\nu_{R} ~,
\ee
This changes eq.(\ref{mnu0}) and $V^0_{PMNS}$ to
\be \label{mnur}
{\cal M}^0_\nu= -VD_D^0\hat{V}D_R^{0~-1}\hat{V}^TD_D^{0}V^T~, \ee
\be V^0_{PMNS}=V^{T}V{\cal F}^0 ~, \ee
where $\hat{V}=V^{T} \tilde{V}_R$ is a general unitary matrix and ${\cal F}^0$ is defined 
as
\be 
\label{f0} 
{\cal F}^{0\dagger} D^0_D\hat{V}D_R^{0~-1}\hat{V}^T D^0_D{\cal F}^{0~*}=D^0_{\nu} ~. 
\ee
Also in this case,  one does get  different mixing matrices for quarks and leptons as 
required. The $V^0_{PMNS}$ is no longer strictly symmetric. 
Corrections due to ${\cal F}^0$ can be small due to hierarchy 
in masses $D_D^0$  if the diagonal elements of $\hat{V}D_R^{0~-1}\hat{V}^T$ are not 
suppressed due to phases in $\hat{V}$. In the converse case, some large mixing angle
may come from ${\cal F}^0$  and the restrictions on $\phi,\th$ may get non-trivially 
modified. 

The realization of the second version of our ansatz would require special
neutrino structures and some additional symmetry to realize it.  A simple 
possibility is using the type-II seesaw \cite{seesaw2} in which case one requires
eq. (\ref{type2}). The same version can be realized 
using type-I seesaw provided that $M_R$ has a specific structure. 
 Assuming that all the Dirac mass matrices, including the Dirac mass 
matrix of 
neutrinos have the same universal structure as 
(\ref{mat-f}) we can determine  $M_R$ to have a form
\beqa \label{t2}
M_R &=& V D_D V_{PMNS}^0 D_\nu^{-1} V_{PMNS}^0 D_D V^T ~,\nonumber \\
&=&(VD_DV^T)(VD_\nu^{-1}V^T)(VD_DV^T)~.    
\eeqa
All the matrices should be non-singular here.  Notice that $M_R$
turns out be product of three diagonal mass matrices all rotated through $V$.
It would be interesting to look for special symmetries enforcing 
eq. (\ref{t2})

\newpage
\noindent Before closing, let us recapitulate:

\noindent
1).  Motivated by the SU(5) relation between mass matrices of the 
charged leptons and down quarks we proposed a universal ansatz for the fermion 
mixing. According to this ansatz  there exists certain
(universality) basis in which   mass matrices of all fermions but charge leptons (or 
neutrinos)  are diagonalized by the same rotation matrix $V$ at the  lowest order. The matrix for charged 
leptons (or neutrinos) is diagonalized by the conjugate $V^*$.

\noindent This leads to an absence of quark mixing but 
non-trivial lepton mixing given by $V_{PMNS}^0 =  V^T V$. 

\noindent In this way we have realized an idea that 
difference in the quark and lepton mixing  appears 
in some  lowest approximation: the quark mixing is zero 
and the lepton mixing is non-trivial and, in general, large. 
The CKM mixing is realized as the correction. \\

\noindent
2).  We studied properties of the lowest order  mixing matrix.  
$V_{PMNS}^0$ is symmetric matrix characterized by two physical 
angles $\theta$ and $\phi$. We showed that $V_{PMNS}^0$ is rather 
close to the observed mixing matrix for  
$\theta/2 \approx \phi = 20 - 25^{\circ}$. Furthermore, the generic feature 
of $V_{PMNS}^0$ is that the 1-3 mixing is close to the present upper bound. \\

\noindent
3). The universal mixing matrix can follow 
from universal mass matrices of fermions. 
A natural realization is the one in which  
the mass matrices have only one (third) non zero eigenvalue
in the lowest order. \\

\noindent
4).  We considered corrections to the lowest (universal) approximation 
which (a) generate the CKM-mixing; 
(b) modify $V_{PMNS}^0$ leading to better agreement with  
observations, in particular, decrease $\sin \theta_{13}$.

Assuming (for simplicity) the  quark - lepton universality,  
in which the  correction  matrix for lepton mixing  equals 
(or similar to) the CKM matrix we obtain good agreement of 
$V_{PMNS}^0$ with data. In this case we predict 
$\sin \theta_{13} > 0.08$ which can be tested in the next generation of experiments. \\

\noindent
5).  The suggested ansatz can be embedded  into 
SU(5) and SO(10) GU models. \\

\end{document}